\documentclass[%
 reprint,
superscriptaddress,
 amsmath,amssymb,
 aps,prl
]{revtex4-1}

\usepackage{graphicx}
\usepackage{dcolumn}
\usepackage{bm}
\usepackage{ulem} 

\usepackage[usenames, dvipsnames]{color}

\begin{document}

\preprint{APS/123-QED}

\title{Transition in swimming direction in a model self-propelled inertial swimmer}

\author{Thomas Dombrowski}
\thanks{Contributed equally}
\affiliation{%
Department of Physics, The University of North Carolina at Chapel Hill, Chapel Hill, NC, USA}

\author{Shannon K. Jones}
\thanks{Contributed equally}
\affiliation{%
Department of Applied Physical Sciences, The University of North Carolina at Chapel Hill, Chapel Hill, NC, USA
}%

\author{Georgios Katsikis} 
\affiliation{Koch Institute for Integrative Cancer Research, Massachusetts Institute of Technology, Cambridge, Massachusetts, USA}%

\author{Amneet Pal Singh Bhalla}%
\affiliation{Applied Numerical Algorithms Group, Lawrence Berkeley National Laboratory, Berkeley, CA, USA}%
\affiliation{Mechanical Engineering, San Diego State University, CA, USA}

\author{Boyce E. Griffith} 
\affiliation{Departments of Mathematics, Applied Physical Sciences, and Biomedical Engineering, The University of North Carolina at Chapel Hill, Chapel Hill, NC, USA}%

\author{Daphne Klotsa} 
\email{dklotsa@email.unc.edu}
\affiliation{%
Department of Applied Physical Sciences, The University of North Carolina at Chapel Hill, Chapel Hill, NC, USA
}%

\date{\today}

\begin{abstract}

We propose a reciprocal, self-propelled model swimmer at intermediate Reynolds numbers ($Re$). Our swimmer consists of two unequal spheres that oscillate in antiphase generating nonlinear steady streaming (SS) flows. We show computationally that the SS flows enable the swimmer to propel itself, and also switch direction as $Re$ increases. We quantify the transition in the swimming direction by collapsing our data on a critical $Re$ and show that the transition in swimming directions corresponds to the reversal of the SS flows. Based on our findings, we propose that SS can be an important physical mechanism for motility at intermediate $Re$.

\begin{description}
\item[PACS numbers]
May be entered using the \verb+\pacs{#1}+ command.
\end{description}
\end{abstract}

\pacs{Valid PACS appear here}
\maketitle

Understanding  motility requires connections between fundamental physics  and biology~\cite{Nachtigall2001,vogel1988life,Vogel1994} and has many applications, including drug-delivering nanomachines~\cite{wang2015one,patra2013intelligent} and autonomous underwater vehicles~\cite{tolba2015taking,fujiwara2014self,duarte2016evolution}.
Swimming regimes can be classified by the Reynolds number ($Re$), which characterizes the relative importance of inertial over viscous forces. Although there is a large body of work on motility in Stokes flows ($Re\ll 1$), in which viscous forces dominate, and at high $Re$ ($Re\gg 1$), in which inertial forces dominate, less is known about the intermediate regime $Re_\text{int}\sim 1\text{ -- }1000$~\cite{childress-book,vogel1988life,Lauga2009}.

The $Re_\text{int}$ regime encompasses an enormous diversity of organisms, ranging from larvae (of \textit{e.g.} fish, squid, ascidian) and large ciliates, to nematodes, copepods, plankton and jellyfish, that exhibit a variety of motility mechanisms: jet propulsion~\cite{Bartol2009,Herschlag2011}, anguilliform locomotion ~\cite{Kern2006,fuiman1988ontogeny,Sznitman2010,McHenry2003,Bhalla13FD}, rowing~\cite{Strickler1975,Blake1986}, aquatic flapping flight~\cite{Borrell2005}, and ciliate beating~\cite{gemmell2015tale,jiang2011does}. Plankton have even been proposed to contribute to the large-scale transport of nutrients and dissolved gases in the ocean~\cite{kunze2006observations,wilhelmus2014observations,Nawroth2014,houghton2018vertically,chisholm2018partial}. However, most prior studies on $Re_\text{int}$ motility have focused on the details of specific organisms ~\cite{Bartol2009,Herschlag2011,Kern2006,fuiman1988ontogeny,McHenry2003,Bhalla13FD,Strickler1975,Blake1986,Borrell2005,gemmell2015tale,jiang2011does,wilhelmus2014observations,Nawroth2014,jones2016bristles}. As a result, few general models exist for motility at $Re_{int}$; examples are an extension of the Stokesian squirmer to include inertia~\cite{lauga-continuous,arezoodaki_2012inertial,khair2014expansions,chisholm16,Li2016,chisholm2018partial}, which makes assumptions about the generation of flow due to small-amplitude oscillations on the surface of a spherical swimmer and the flapping-plate model, which is a lumped-torsional-flexibility model that uses passive pitching and responds to an actuation~\cite{zhang2010locomotion,spagnolie2010surprising}. However, there is a lack of understanding regarding the unifying physical mechanisms that swimmers at $Re_\text{int}$ exhibit. To achieve this, more models with varying degrees of freedom that operate under different conditions at $Re_\text{int}$ are needed. Only then can we make progress in better understanding biological swimmers and designing artificial ones at intermediate scales.

Steady streaming (SS) is the nonzero, time-averaged flow that arises at $Re_\text{int}$ due to oscillations of a rigid body in a fluid and has been studied for various cases, such as around a single sphere~\cite{riley66,riley2001,chang1994unsteady,kotas2007visualization,otto2008measurements}, cylinder, near a wall. 
While SS has been used to manipulate particles \textit{e.g.} \cite{klotsa2007,wright2008,klotsa2009,klotsa2015propulsion,spelman2017} and cells~\cite{lutz2006hydrodynamic} via \textit{external} vibrations, it has not been used as a mechanism for self-propulsion, even though there have been suggestions that it may be relevant for the enhanced motility of \textit{Synechococcus} cyanobacteria~\cite{ehlers2011}.

\begin{figure}
\includegraphics[width=1\columnwidth]{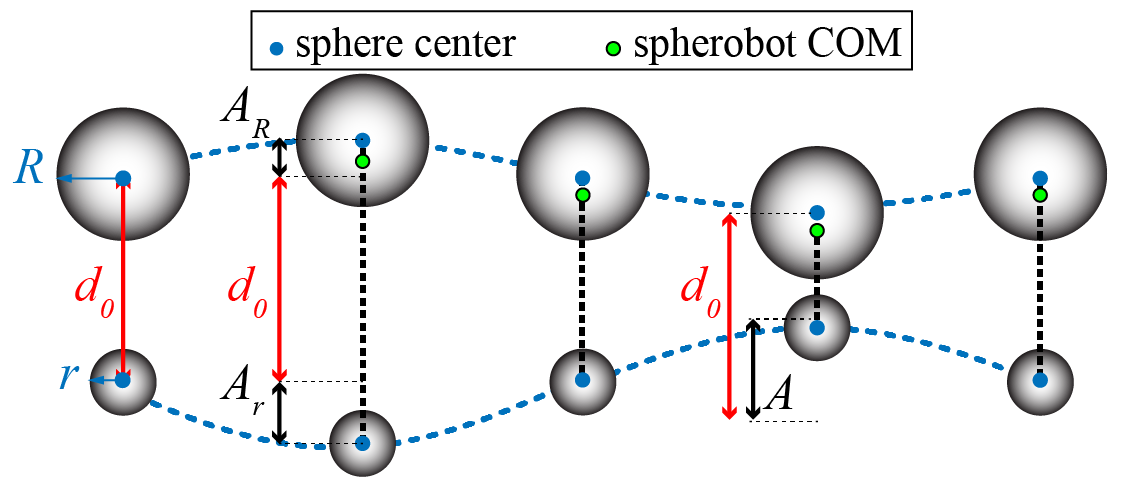}
\centering
\caption{Reciprocal oscillation of the spherobot swimmer over one cycle. Spheres' centers of mass (COM)(blue circles) and the spherobot COM (green circle) are indicated. The distance between the spheres' centers, $d(t)$, is $d_0$, at the equilibrium distance, $d_0-A$, minimum distance and $d_0+A$ at maximum distance. The total amplitude $A=A_R+A_r$. 
}
\label{diagram}
\end{figure}

In this letter we propose a simple, reciprocal, and self-propelled model swimmer, termed the \textit{spherobot}, that uses steady streaming flows in a novel way, \textit{i.e.} for propulsion. The spherobot is composed of two unequal spheres that oscillate with respect to each other, in antiphase, generating SS flows, Fig.~\ref{diagram}(a). 
We computationally studied the spherobot's motility over a broad range of parameters: viscosity, sphere amplitudes, distance between the spheres, sphere radii and sphere-radii ratio. At $Re=0$, the spherobot cannot swim because of Purcell's scallop theorem~\cite{Purcell1977}; its reciprocal stroke does not break time-reversal symmetry.
At low, nonzero $Re$ the spherobot started to swim and, interestingly, switched swimming direction from a small-sphere-leading to a large-sphere-leading regime. We found that the point of transition collapsed to a critical value when the appropriate Reynolds number was used, which revealed a strong dependence on the SS flows of the small sphere. Analyzing the flow fields, we showed that the transition in swimming direction corresponds to the reversal of SS flows around the spherobot that occurs as the Reynolds number increases.

\textbf{\textit{Methods.}}
The spherobot was composed of two unequal spheres with radii $r$, $R$, which were coupled to one another by prescribing the distance between their centers. To model this computationally, we tethered the two spheres using an active spring with a time-dependent distance $d(t)=d_0+A \sin(\omega t)$, in which $d_0$ is the equilibrium distance between the sphere centers, $A=0.5(d_\text{max}-d_\text{min})$ is the amplitude of the spherobot, and $\omega$ is the frequency of oscillation (Fig.~\ref{diagram}). Equal and opposite (spring) forces were applied to the spheres that acted to keep them approximately at the prescribed distance apart (error $\approx 10^{-7}$m). Thus, the model ensures a geometrically reciprocal cycle and a force-free swimmer. Because the same force is applied to both spheres, the one with the smaller mass (the small sphere) will have a larger amplitude $A_r$ than the one with the bigger mass (large sphere), $A_R$, (\textit{i.e.} if $r \leq R$ then $A_r \geq A_R$). In most simulations we have $A_r \approx 4 A_R$. The amplitude of the spherobot $A$ is the sum of the two, $A$ $=A_r+A_R$, Fig.1. Both spheres were neutrally buoyant with respect to the surrounding fluid. 
To simulate the spherobot in a fluid, we used an exactly constrained immersed boundary (CIB) method~\cite{Kallemov16,Usabiaga17}.  The CIB scheme is implemented in IBAMR ~\cite{griffith2007adaptive,IBAMR-web-page}, which provides several variants of the immersed boundary (IB) method~\cite{Peskin2002} for fluid-structure interaction. The spherobot was immersed in a fluid that occupied a finite cell with no-slip walls. The visualization and analysis of the flow fields was done in VisIt~\cite{HPV:VisIt}. Further details on the model and
method are given in the Supplemental Material (SM)~\cite{SM}.

The swimming velocity of the spherobot was measured after steady state had been reached and was defined as the net displacement of the spherobot center of mass over one cycle.
We defined the Reynolds number as $Re=A_rr/\delta^2=A_rr\omega/\nu$ because, as we will show, it is the ratio that determines the transition between small-sphere-leading and large-sphere-leading regimes; $\delta=\sqrt{\nu/\omega}$ is the oscillatory boundary layer thickness and $\nu$ is the kinematic viscosity of the fluid. We carried out simulations in 2D and 3D and found qualitative agreement. We focused on 2D because it allowed us to study a much broader parameter space. The range studied in $Re$ was $0.001\le Re \le 150$. All other parameter ranges (amplitude, radii,\textit{etc.}) are shown in the SM, table S1~\cite{SM}.


\begin{figure}
\includegraphics[width=1\columnwidth]{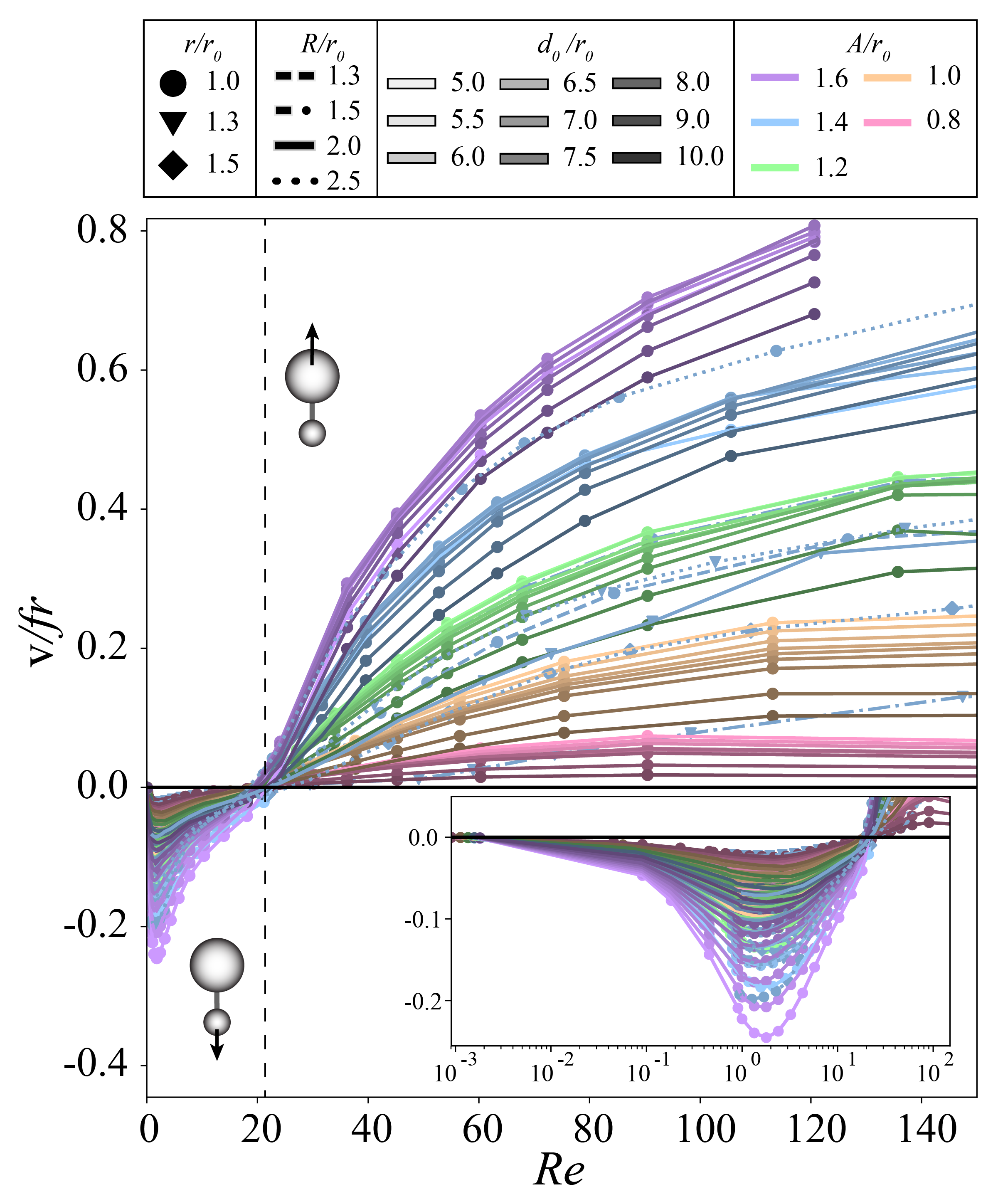}
\centering
\caption{Velocity of the spherobot as a function of $Re$ in 2D for a range of $A$, $d_0$, $R$, $r$ shown in the legend. The inset shows the small-sphere-leading regime plotted on a semi-log-x scale. Parameters $A$, $d_0$, $R$, and $r$ are non-dimensionalized by the length scale, $r_0=0.15$m, the radius for the small sphere. Line color indicates $A$, line saturation indicates $d_0$, line style indicates $R$, and symbols indicate $r$. Negative velocity indicates swimming in the direction of the small sphere, positive velocity indicates swimming in the direction of the large sphere. 
Vertical dashed lines denote critical $Re$ for transition.
}
\label{VvsRe}
\end{figure}

\begin{figure*}
\includegraphics[width=2\columnwidth]{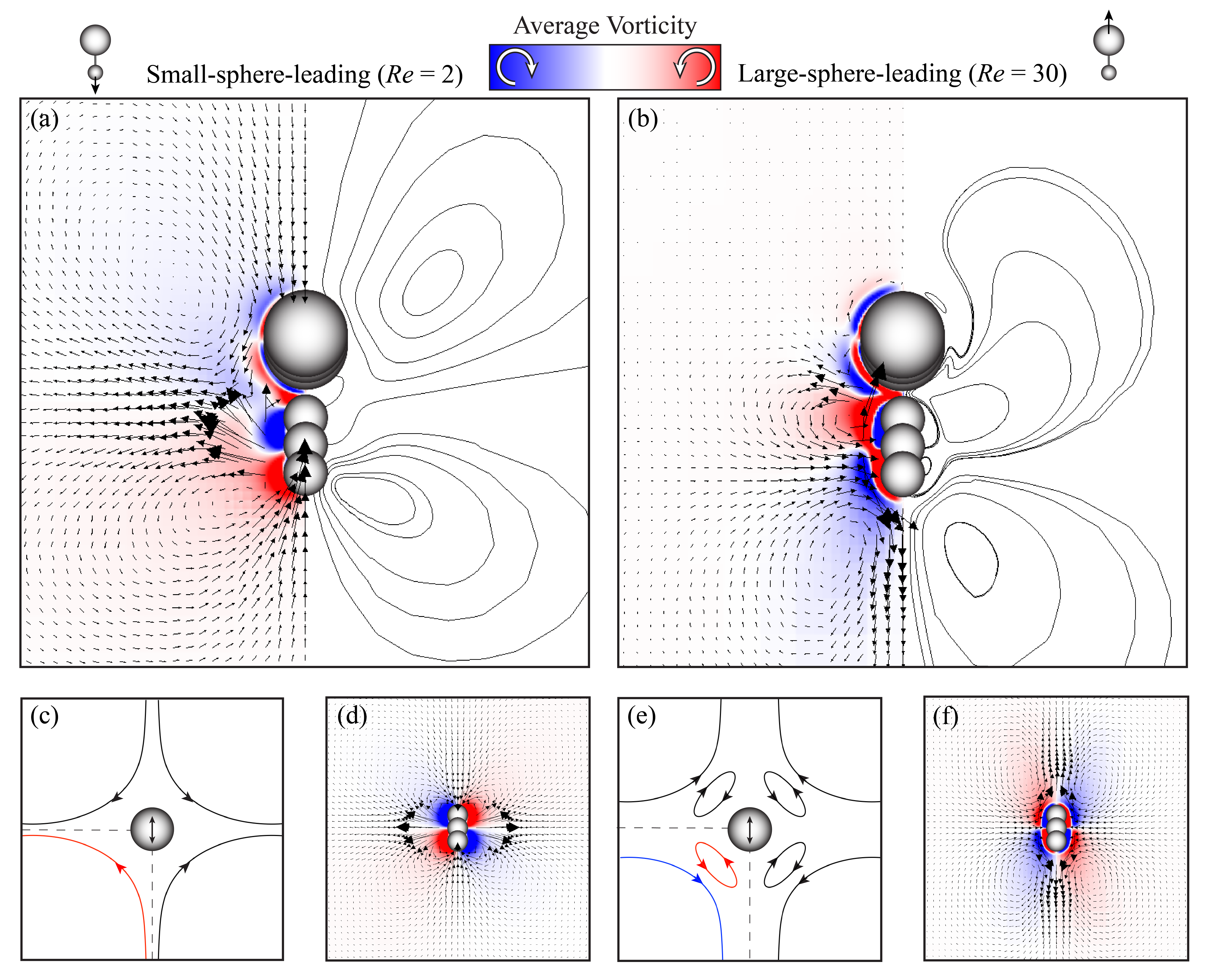}
\centering
\caption{Left column: Small-sphere-leading regime at $Re$ = 2. Right column: Large-sphere-leading regime at $Re$ = 30. Spherobot velocity field superimposed with the vorticity field and streamlines in (a) the small-sphere-leading regime and (b) the large-sphere-leading regime. The largest dimensionless velocity magnitude in (a), (b) is $|\mathbf{v_{max}}|/fr = 0.88$. Schematic diagrams showing the reversal of steady streaming flows for one sphere in the limiting cases (c) $\delta\gg r$, and (e) $\delta\ll r$. Due to symmetry the lower left quadrant is indicated with a dashed line. Velocity vector plot superimposed with the vorticity field for one sphere at (d) $Re=2$ and (f) $Re=30$. The largest dimensionless velocity magnitude in (d), (f) is $|\mathbf{v_{max}}|/fr = 1.1$. All velocity vectors are scaled the same.}
\label{FlowStream}
\end{figure*}

\textbf{\textit{Results.}} 
We initially placed the spherobot in the simulation box at constant $A$, $A_R$, $A_r$, $d_0$, $f$, $R$, $r$ and varied the $Re$ via the kinematic viscosity, $\nu$. As a validation, we ran a simulation at $Re=0$ and confirmed that the spherobot did not swim because of Purcell's theorem for reciprocal swimmers~\cite{Purcell1977}. As soon as $Re>0$ (lowest value $Re=0.001$) the spherobot began to swim in the direction of the small sphere (Fig.\ref{VvsRe}), \textit{i.e.} the small-sphere-leading regime. As $Re$ increased, 
the speed of the spherobot increased until reaching a maximum at $Re\approx 2$. Above $Re\approx 2$ the spherobot slowed down and eventually had no net displacement (even though the spheres oscillated) at $Re\approx 20$. As $Re$ increased further, the spherobot switched direction to swim with the large sphere on the front, \textit{i.e.} the large-sphere-leading regime, where its increasing speed started to plateau as $Re$ increased further.
We then ran a broader parameter sweep varying $R$, $r$, $A$, $A_R$, $A_r$, and $d_0$ besides the viscosity $\nu$. We found that the transition only depended on the small sphere's radius and amplitude (besides viscosity) and that it was independent of all the other length scales $R$, $A_R$ and $d_0$. The transition-point data collapsed (within the scatter on a single, critical dimensionless number $Re=A_rr/\delta^2\approx 20$ (Fig.\ref{VvsRe}). 

 To gain insight into the propulsion mechanism and the switch in swimming direction, we turned our attention to the flow fields generated by the spherobot. Based on classical work on steady streaming generated by a single oscillating sphere, we expected each sphere of the spherobot to generate SS flows, which are time-averaged flows by definition. We also anticipated the SS flows around the spherobot to be different than the classical SS flows around a sphere for two reasons. First, the small sphere's oscillation amplitude, $A_r$, was of the same order of magnitude as the sphere radius, \textit{i.e.} $\epsilon=A_r/r \approx O(1)$ unlike the assumption for classical steady streaming where $\epsilon\ll 1$~\cite{andres1953acoustic,riley66,riley2001}. Second, it was unclear what the cumulative SS flows of two spheres oscillating in antiphase should be, as it has only been studied for spheres and cylinders in phase~\cite{zapryanov1988Cylinders,tabakova1982AxiSpheres,tabakova19823DSpheres,klotsa2007}.
 Bearing these considerations in mind, we calculated the time-averaged flow fields around the spherobot, varying the same parameters as before, (Fig.2). We found that the switch in the swimming direction at $Re\approx20$ corresponded to the reversal of the SS flows both parallel and perpendicular to the axis of oscillation.
Specifically, in the small-sphere-leading regime ($Re<20$), the fluid, on average, was pulled in towards the spheres along the axis of oscillation and was pushed out away from the spheres along the axis perpendicular to the oscillation, (Fig.~\ref{FlowStream}~(a)). On the contrary, in the large-sphere-leading regime ($Re>20$), the fluid, on average, did the opposite -- it was pushed away from the spheres along the direction of swimming (with a strong downward jet below the small sphere) and was pulled in towards the gap between the spheres in the direction perpendicular-to-swimming (Fig.~\ref{FlowStream}~(b)). 
 
 
 Furthermore, in both regimes it is clear that the velocity vectors along the oscillation axis are larger around the small sphere than the large sphere, (Fig.3). In fact, through control volume analysis we found that for both regimes, the momentum flux on the side of the small sphere was larger than the momentum flux on the side of the large sphere, (the ones along the perpendicular axis generated fluxes that canceled each other). Though initially unexpected, this finding makes sense together with the collapse, which depends on the $Re$ of the small sphere only, (Fig.2). The net momentum flux of course switches direction as the swimming direction switches, (see Fig.S4~\cite{SM}). Our data, thus, strongly suggest that the transition in the spherobot's swimming direction is due to the reversal of SS flows, which is associated with the switch in the direction of the net momentum flux~\cite{SM}.

\textbf{\textit{Discussion}}. To better understand the reversal of SS flows, we will consider what is known for one sphere.  
Analytic solutions have been obtained under the small-amplitude assumption $A_r\ll r$ and in the two limiting cases relating the sphere radius to the boundary layer thickness, $\delta \gg r$ and $\delta \ll r$. The two limiting cases demonstrate a reversal in direction, shown schematically in Fig.3(c),(e)~\cite{riley66,riley2001,andres1953acoustic}. In the first case, the boundary layer thickness is much larger than the radius, $\delta \gg r$, (Fig.3(c)). Due to symmetry we describe one quadrant of flow. A single vortex that is the boundary layer is generated near the surface of the sphere, which pulls fluid along the axis of oscillation and pushes fluid out in the perpendicular. In the second case, the boundary layer thickness is much smaller than the radius of the sphere, $\delta \ll r$, (Fig.3(e)). Two vortices are generated swirling in opposite directions. The boundary layer is confined into an inner vortex close to the surface of the sphere (same direction as in the first case) but there is an additional outer vortex in the opposite direction -- it pushes fluid out along the axis of oscillation and pulls it in along the perpendicular. The analytical limiting solutions, that we just described, provide us a with qualitative picture; we cannot use them for direct comparison because neither $A_r\ll r$ nor $\delta \gg r$ or $\delta \ll r$ holds true for our system. Instead, we compare our results to experiments and simulations, where $\epsilon=A_r/r=O(1)$ and $r/\delta=O(1)$, as for the spherobot. 


Unlike the spherobot where the reversal of flows corresponds to a switch in the direction of swimming, the point where reversal of flows occurs for a single sphere is not well-defined~\cite{riley66,chang1994unsteady,Alassar1997,blackburn2002}. Experimental observations have reported that they could observe the inner vortex when $r/\delta \leq 4$~\cite{Tatsuno1973LowRe}, while experiments and simulations reported the coexistence of inner and outer vortices with opposing flows for $r/\delta \geq 7$~\cite{coenen2008Reversal,Tatsuno1973LowRe,tatsuno1981Unharmonic,olsen1956rotational,holtsmark1954boundary}. It was also shown that the reversal of flows depends on the sphere's amplitude, yet a specific scaling was not found~\cite{Tatsuno1973LowRe,chang1994unsteady,Alassar1997,blackburn2002}. Relating all this back to the spherobot, our data collapse gave $Re=A_r r/\delta^2$ as the critical parameter for the transition in swimming direction, a scaling that includes an amplitude dependence, as suggested by previous works. 
Moreover, we found that when plotting the dimensionless velocity of the spherobot as a function of $r/\delta$, the transition in swimming ($Re\approx 20$) occurred in the range $r/\delta \approx [3.5,7]$, (see Fig.S7~\cite{SM}), again in agreement with previous reports on the reversal of SS flows for a sphere.

 We can make an analogy that the large sphere of the spherobot acts like the body of the swimmer while the small sphere acts like the flagellum. In fact, it is really interesting that the SS flows, which are unrelated to the squirmer models, in the small-sphere leading regime resemble the flow field of Stokesian pullers and in the large-sphere leading regime resemble the flow field of Stokesian pushers~\cite{lighthill1952squirming,blake1971,Pedley16}. However, the organisms that swim like pullers and pushers such as algae and bacteria, respectively, have different appendages in order to perform the ``pulling'' or the ``pushing''. What is remarkable here is that the geometry of the spherobot does not have to change -- the small sphere can act as an effective flagellum that can both ``pull'' and ``push'' depending only on the critical parameter $Re=A_r r/\delta^2$. \textit{E.g.} our swimmer can change its amplitude and it will switch swimming direction.
 
 

To conclude, we have proposed a model spherobot swimmer that utilizes SS in a novel way, to propel itself. The main findings of the current letter are (i) a transition in the swimming direction that collapses onto a single critical Reynolds number and (ii) the physical mechanism for the transition in swimming is the reversal of SS flows. Based on our findings, we propose that SS can be an important physical mechanism present more generally in motility at $Re_\text{int}$ both in biological organisms but also when designing artificial swimmers~\cite{ehlers-arxiv,ehlers2011,nadallauga2014asymmetric,spelman2017}.  
Finally, we expect to find interesting emergent collective behavior of multiple spherobot swimmers as nonlinearities add up leading to different steady states and patterns. 



\acknowledgments
\textbf{Acknowledgments.}
S.K.J. and T.D. contributed equally to this work.
D.K. acknowledges the National Science Foundation, grant award DMR-1753148.

\clearpage
\bibliography{bibliography}

\end{document}